\documentclass[aps,prb,twocolumn,showpacs,amsmath,amssymb,superscriptaddress,longbibliography]{revtex4-1}

\usepackage{bm}
\usepackage{amsthm}
\usepackage{amssymb}
\usepackage{amsfonts}
\usepackage{graphicx}
\usepackage{epsfig}
\usepackage{subfigure}
\usepackage[colorlinks=true,citecolor=blue,urlcolor=blue]{hyperref}
\usepackage{color}
\usepackage{ulem}

\begin{document}

\title{Surface energy of the one-dimensional supersymmetric $t-J$ model with general integrable boundary terms in the antiferromagnetic sector}
\author{Pei Sun}
\affiliation{School of Physics, Northwest University, Xi'an 710127, China}
\affiliation{Shaanxi Key Laboratory for Theoretical Physics Frontiers, Xi'an 710127, China}
\author{Yang-Yang Chen}
\affiliation{Shaanxi Key Laboratory for Theoretical Physics Frontiers, Xi'an 710127, China}
\affiliation{Institute of Modern Physics, Northwest University, Xi'an 710127, China}
\author{Tao Yang}
\email{yangt@nwu.edu.cn}
\affiliation{School of Physics, Northwest University, Xi'an 710127, China}
\affiliation{Shaanxi Key Laboratory for Theoretical Physics Frontiers, Xi'an 710127, China}
\affiliation{Peng Huanwu Center for Fundamental Theory,  Xi'an 710127, China}
\affiliation{Institute of Modern Physics, Northwest University, Xi'an 710127, China}
\author{Junpeng Cao}
\email{junpengcao@iphy.ac.cn}
\affiliation{Beijing National Laboratory for Condensed Matter Physics, Institute of Physics, Chinese Academy of Sciences, Beijing 100190, China}
\affiliation{Peng Huanwu Center for Fundamental Theory,  Xi'an 710127, China}
\affiliation{Songshan Lake Materials Laboratory, Dongguan, Guangdong 523808, China}
\affiliation{School of Physical Science, University of Chinese Academy of Sciences, Beijing 100049, China}
\author{Wen-Li Yang}
\email{wlyang@nwu.edu.cn}
\affiliation{School of Physics, Northwest University, Xi'an 710127, China}
\affiliation{Shaanxi Key Laboratory for Theoretical Physics Frontiers, Xi'an 710127, China}
\affiliation{Peng Huanwu Center for Fundamental Theory,  Xi'an 710127, China}
\affiliation{Institute of Modern Physics, Northwest University, Xi'an 710127, China}
\date{\today}


\begin{abstract}
In this paper, we study the surface energy of the one-dimensional supersymmetric $t-J$ model with unparallel boundary magnetic fields, which is a typical
$U(1)$-symmetry broken quantum integrable strongly correlated electron system. It is shown that at the ground state, the contribution of inhomogeneous term in the Bethe ansatz solution of eigenvalues of transfer matrix satisfies the finite size scaling law $L^{\beta}$ where $\beta<0$.
Based on it, the physical quantities of the system in the thermodynamic limit are calculated. We obtain the patterns of Bethe roots and the analytical expressions of density of states, ground state energy and surface energy.
We also find that there exist the stable boundary bound states if the boundary fields satisfy some constraints.

\end{abstract}


\maketitle


\section{Introduction}
\label{sec1} \setcounter{equation}{0}

The $t-J$ model plays important roles in the strongly correlated electronic systems especially
for characterizing the high-$T_c$ superconductivity \cite{hubbard01, zhang1988, zeng02, hubbard02, hubbard04}.
The model Hamiltonian includes the nearest neighbor hopping ($t$) and the antiferromagnetic exchanging interaction ($J$).
The $t-J$ model can be obtained by taking the large on-site Coulomb repulsion that excludes the
double-occupancy of every site of the Hubbard model.

At the point of $J=\pm 2t$, the one dimensional $t-J$ model is supersymmetric and can be solved exactly.
By using the coordinate Bethe ansatz, Lai \cite{Lai} and Sutherland \cite{zeng06} obtain the exact solution and Bethe ansatz equations of the system with periodic boundary condition.
Later, it is found that these two solutions are equivalent by introducing a particle-hole transformation \cite{sarkar1990}.
Based on the quantum group invariant, the graded nested algebraic Bethe asnatz method is proposed and applied to the one dimensional supersymmetric $t-J$ model \cite{zeng03,essler199202,zeng04,zeng05,zeng08}.

The next task is to solve the obtained Bethe ansatz equations. Schlottmann obtained the ground state Bethe roots distribution of the system in an external magnetic field
and calculate the thermodynamic quantities such as free energy and magnetic susceptibility \cite{schlo1987}.
Bares et al calculate the exact ground state and excitation spectrum of the system \cite{Bares-1991}.
Other interesting properties of the system such as charge-spin separation \cite{bares1990}, correlation functions \cite{1}
finite-temperature thermodynamics and excitations \cite{2}, Luttinger liquid behavior \cite{3}
and crossover phenomena in the correlation lengths \cite{zeng10} are also studied.

Besides the periodic boundary condition, the integrable open one is another typical quantization conditions \cite{zeng01}.
The boundary reflection of electrons can be tuned by the boundary magnetic fields.
If the boundary fields are parallel, the traditional coordinate and algebraic Bethe ansatz still work.
Then the exact physical properties of the system has be studied extensively \cite{zeng09,essler1996,zeng11,4,5,zeng101,zeng108,hubbard03}.
If the boundary fields are not parallel, which breaks the $U(1)$ symmetry of the system but the integrability still holds,
it is very hard to find the suitable vacuum states and the traditional Bethe ansatz does not work. Then the off-diagonal Bethe asnatz is proposed \cite{Cao13,wang2015}.
The eigenvalue of the transfer matrix is characterized by the inhomogeneous $T-Q$ relations \cite{zhang2014}.
We should note that due to the existence of inhomogeneous terms, it is
hard to study the physical quantities in the thermodynamic limit because the associated Bethe ansatz equations are not in the form of product and
the usual thermodynamic Bethe ansatz \cite{takasha} can not be applied.
The patterns of roots of inhomogeneous Bethe ansatz equations are very complicated.
Only for some special cases, the distribution of roots of the degenerate Bethe ansatz equations are found and the related
physical properties are studied \cite{wen2018}.

In this paper, we investigate the physical quantities in the thermodynamic limit of the one-dimensional supersymmetric $t-J$ model with unparallel boundary fields. The Hamiltonian reads
\begin{eqnarray}
\label{Hamiltonian}
  H &=& -t\sum_{j=1}^{L-1}\sum_{\sigma=\uparrow, \downarrow} \mathbb{P}\left[c_{j,\sigma}^\dag c_{j+1,\sigma}+
  c_{j+1,\sigma}^\dag c_{j,\sigma}\right]\mathbb{P}\nonumber\\
  &&+J\sum_{j=1}^{L-1}\left[\boldsymbol{S}_j \cdot \boldsymbol{S}_{j+1}-\frac{1}{4}n_jn_{j+1}\right]\nonumber\\
  &&+\chi_1n_1+2\boldsymbol{h}_1 \cdot \boldsymbol{S}_1+\chi_Ln_L+2\boldsymbol{h}_L\cdot \boldsymbol{S}_L.
\end{eqnarray}
Here $c_{j,\sigma}^\dag $ and $c_{j,\sigma}$ are the creation and annihilation operators of electrons with spin $\sigma=\uparrow, \downarrow$ at $j$-th site, respectively. $L$ is the length of system size.
$n_j=\sum_{\sigma}n_{j\sigma}$ and $n_{j\sigma}=c_{j,\sigma}^\dag c_{j,\sigma}$ are the electron number operators.
$\boldsymbol{S}_j$ is the spin operator of $j$-th electron. By using the creation and annihilation operators, the components of spin operator can be written as
$S_j=c_{j,\uparrow}^\dag c_{j,\downarrow}$, $S_j^\dag =c_{j,\downarrow}^\dag c_{j,\uparrow}$ and $S_j^z=\frac{1}{2}
(n_{j,\uparrow}-n_{j,\downarrow})$, which are the generators of $su(2)$ symmetry in the spin sector.
$\hat{N}=\sum_{j=1}^L n_j$ is the total number of electrons.
$\mathbb{P}=\Pi_{j=1}^{L}(1-n_{j\uparrow}n_{j\downarrow})$ is the projector which is included to ensure the constraint of no double occupancy.
$\boldsymbol{h}_1=(h_1^x, h_1^y, h_1^z)$ and $\boldsymbol{h}_L=(h_L^x, h_L^y, h_L^z)$
are the unparallel boundary magnetic fields. $\xi_1$ and $\xi_L$ are the boundary chemical potentials.
Without losing generality, we put $t=1$. In this paper, we focus attention on the sector of $J=2t=2$.

We propose a method to calculate the physical quantities induced by the unparallel boundary fields.
The main idea is as follows.
From the finite size scaling analysis, we find that at the ground state,
the inhomogeneous term in the eigenvalue of transfer matrix, which is the generating function of all the conserved quantities of the system including the Hamiltonian \eqref{Hamiltonian},
can be neglected in the thermodynamic limit. Based on it, we obtain the patterns of roots of the reduced Bethe ansatz equations.
Thus, when the system size tends to infinity, the integral equation of the density of states is achieved.
We find that there exist the pure imaginary Bethe roots which correspond to the stable boundary bound states, if the boundary fields satisfy some constraints.
We obtain the analytical expressions of the ground state energy with arbitrary filling factor and the surface energy with half filling.
In order to check the correction of obtained results, we also calculate these physical quantities by the density matrix renormalization group (DMRG) and the
finite size scaling analysis. The analytical results are consistent with the numerical ones very well.

The paper is organized as follows. In section \ref{sec2}, we introduce the off-diagonal Bethe ansatz solutions of the model \eqref{Hamiltonian}.
In section \ref{sec3}, we study the finite size scaling behavior of the ground state energy. In section \ref{sec3.0},
we calculate the ground state energy and the surface energy induced by the boundary fields in the thermodynamic limit.
We summarize the results and give some discussions in section \ref{sec6}.

\section{Bethe ansatz solutions}
\label{sec2}

The integrability of the model (\ref{Hamiltonian}) is associated with the graded $9\times 9$ $R$-matrix
\begin{equation}\label{R}
R_{0,j}(u)=u+\Pi_{0,j},
\end{equation}
where $u$ is the spectral parameter and $\Pi_{0,j}$ is the $Z_2$-graded permutation operator.
The $R$-matrix \eqref{R} is defined in graded tensor space $V_0\otimes_s V_j$,
where $V_0$ is the 3-dimensional auxiliary space, $V_j$ is the 3-dimensional quantum or physical spaces and
the notation $\otimes_s$ means the graded tensor. In this paper, we adopt the graded tensor with the definition
$[A\otimes_s B]^{a_2 b_2}_{a_1 b_1}= (-1)^{p_{a_2}p_{b_2}}A_{a_1}^{a_2} B_{b_1}^{b_2}$, where $A$ and $B$ are two arbitrary vectors,
the raw indices $\{a_1, b_1\}$ and column indices $\{a_2, b_2\}$ take the values in the set $\{1, 2, 3\}$, $p_\gamma$ is the Grassmann parity, $p_1=0$ and $p_2=p_3=1$.
Thus $(\Pi_{0,j})^{\gamma\delta}_{\alpha\beta}=(-1)^{p_{\gamma}
p_{\delta}}\delta_{\alpha,\delta}\delta_{\beta,\gamma}$, which endows the fundamental representation of $su(1|2)$ algebra.
The $R$-matrix (\ref{R}) satisfies the graded Yang-Baxter equation
\begin{eqnarray}
&&R_{0,0'}(u-v)R_{0,j}(u)R_{0',j}(v)\nonumber \\
&& = R_{0',j}(v)R_{0,j}(u)R_{0,0'}(u-v).\nonumber
\end{eqnarray}

In the system \eqref{Hamiltonian}, the boundary reflection of electrons at one end is characterized by the $3\times 3$ reflection matrix
defined in the auxiliary space as
\begin{eqnarray}
  &&K^-(u)=\left(
                   \begin{array}{ccc}
                     \xi+u & 0 & 0 \\
                     0 & \xi+\cos\theta u & \sin\theta e^{i\varphi}u \\
                     0 & \sin\theta e^{-i\varphi}u & \xi-\cos\theta u
                   \end{array}
                 \right), \label{boundary-matrix}
\end{eqnarray}
where $\xi$, $\theta$ and $\varphi$ are the boundary parameters determined by the boundary magnetic field. The reflection matrix \eqref{boundary-matrix}
satisfies the reflection equation
\begin{eqnarray}
  &&R_{0,0'}(u-v)K_0^-(u)R_{0',0}(u+v)K_{0'}^-(v)\nonumber\\
  &&=K_{0'}^-(v)R_{0,0'}(u+v)K_0^-(u)R_{0',0}(u-v).
\end{eqnarray}
The boundary reflection at the other end is characterized by the dual reflection matrix
\begin{eqnarray}
   && K^+(u)=\left(
                   \begin{array}{ccc}
                     \xi'-u & 0 & 0 \\
                     0 & k_1 & k_2\\
                     0 & k_3 & k_4
                   \end{array}
                 \right), \label{boundary-matrix1}\\
   &&k_1=\xi'-\frac{1}{2}-\frac{1-2u}{2}\cos\theta', \;\; k_2=\frac{2u-1}{2}\sin\theta' e^{i\varphi'}, \nonumber \\
   && k_3=\frac{2u-1}{2}\sin\theta' e^{-i\varphi'}, \;\; k_4=\xi'-\frac{1}{2}+\frac{1-2u}{2}\cos\theta',\nonumber
\end{eqnarray}
where $\xi'$, $\theta'$ and $\varphi'$ are the boundary parameters. The dual reflection matrix \eqref{boundary-matrix1}
satisfy the dual reflection equation
\begin{eqnarray}
  &&R_{0,0'}(u-v)K_0^+(v)R_{0',0}(1-u-v)K_{0'}^+(u)\nonumber\\
  &&=K_{0'}^+(u)R_{0,0'}(1-u-v)K_0^+(v)R_{0',0}(u-v).
\end{eqnarray}
We note that the reflection matrices \eqref{boundary-matrix} and \eqref{boundary-matrix1} have the non-diagonal elements.
The monodromy matrix $T_0(u)$ and the reflecting one $\hat{T}_0(u)$ are constructed by the $R$-matrices as
\begin{eqnarray}
  T_0(u) &=& R_{0,L}(u)R_{0,L-1}(u)\cdots R_{0,1}(u),\nonumber \\
  \hat{T}_0(u) &=& R_{1,0}(u)\cdots R_{L-1,0}(u)R_{L,0}(u).
\end{eqnarray}
The transfer matrix is given by
\begin{equation}
  t(u)=str_0\{K_0^+(u)T_0(u)K_0^-(u)\hat{T}_0(u)\}, \label{boundary-matrix2}
\end{equation}
where $str_0$ is the supertrace with the definition $str_0 C=\sum_{\beta=1}^3 (-1)^{p_{\beta}} C_{\beta}^{\beta}$ and $C$ is a matrix.
The transfer matrix \eqref{boundary-matrix2} is the generating function of conserved quantities of the system \eqref{Hamiltonian}. The
Hamiltonian can be obtained by taking the first order derivative of the logarithm of the transfer matrix $t(u)$
\begin{eqnarray}
  H &=& -\frac{1}{2}\frac{d\ln t(u)}{du}\Big|_{u=0}+\frac{1}{2\xi}
  -\frac{1-2\xi'}{2(1-\xi')}\nonumber\\
  &&-2{\hat{N}}+L-1.\label{boundary-matrix3}
\end{eqnarray}
Comparing Eqs.\eqref{boundary-matrix3} and (\ref{Hamiltonian}), we conclude that the model parameters in the Hamiltonian (\ref{Hamiltonian}) can be expressed in terms of the boundary parameters as
\begin{eqnarray}
  &&\chi_1 = -1+\frac{1}{2\xi},\quad
  h_1^x=-\frac{1}{2\xi}\sin\theta\cos\varphi,\quad\nonumber\\
  &&h_1^y=-\frac{1}{2\xi}\sin\theta\sin\varphi,\quad
  h_1^z=\frac{1}{2\xi}\cos\theta,\nonumber \\
  &&\chi_L=-1+\frac{1}{2(1-\xi')},\quad
  h_L^x=\frac{\sin\theta'\cos\varphi'}{2(1-\xi')},\quad\nonumber\\
  &&h_L^y=\frac{\sin\theta'\sin\varphi'}{2(1-\xi')},\quad
  h_L^z=-\frac{\cos\theta'}{2(1-\xi')}.
  \label{boundary_parameters}
\end{eqnarray}

By using the off-diagonal Bethe ansatz, the eigenvalue $\Lambda(u)$ of the transfer matrix $t(u)$ is given by the inhomogeneous $T-Q$ relation
\begin{eqnarray}
\label{eigenvalue}
&& \Lambda(u) = \omega_3(u)(\xi+u)(u+1)^{2L}\frac{Q(u-1)}{Q(u)}-u^{2L}\bar{a}(u)
  \nonumber\\
  &&\qquad \times \frac{Q(u-1)Q^{(1)}(u+1)}{Q(u)Q^{(1)}(u)}
   -u^{2L}\bar{d}(u)\frac{Q^{(1)}(u-1)}{Q^{(1)}(u)}\nonumber \\
 &&\qquad  +2hu^{2L+1}(u-\frac{1}{2})\frac{Q(u-1)}{Q^{(1)}(u)},
\end{eqnarray}
where the functions $\omega_3(u)$, $\bar{a}(u)$, $\bar{d}(u)$, $Q(u)$, $Q^{(1)}(u)$ and the constant $h$ are
\begin{eqnarray}
\label{constans}
  &&\omega_3(u) = \xi'-u-\frac{1}{2u+1}(2\xi'-1),\nonumber\\
  &&\bar{a}(u) = \frac{u-\frac{1}{2}} {u+\frac{1}{2}}
  (u+\xi')(u+\xi), \;\;\; \bar{d}(u)=(u-\xi')(u-\xi),\nonumber\\
  &&Q(u)=\prod_{k=1}^N(u-\tilde{v}_k)(u+\tilde{v}_k+1),\nonumber \\
  &&Q^{(1)}(u)=\prod_{l=1}^N(u-\tilde{\lambda}_l)(u+\tilde{\lambda}_l),\nonumber\\
  &&h=1-[\cos\theta\cos\theta'+\sin\theta\sin\theta'\cos(\varphi-\varphi')].
  \end{eqnarray}
The parameters $\{\tilde{v}_k | k=1, \cdots, N\}$ and $\{\tilde{\lambda}_l|l=1, \cdots, N\}$ are the Bethe roots.
The last term in Eq.\eqref{eigenvalue} is the inhomogeneous term.
The hermitian of Hamiltonian requires that the boundary parameters
$\xi$, $\theta$, $\phi$, $\xi'$, $\theta'$ and $\phi'$ are real.

From the definition, we know that the eigenvalue $\Lambda(u)$ is a polynomial of $u$. Then the residues of right hand side of \eqref{eigenvalue}
at the poles of $\tilde{v}_k$, $-\tilde{v}_k-1$ and $\pm \tilde{\lambda}_l$ must vanish, which give that the $2N$ Bethe roots
$\{\tilde{\lambda}_l\}$ and $\{\tilde{v}_k\}$ must satisfy the inhomogeneous Bethe ansatz equations (BAEs)
\begin{eqnarray}
  &&\left[\xi'-\tilde{v}_k-\frac{2\xi'-1}{2\tilde{v}_k+1}\right]
  (\xi+\tilde{v}_k)(\tilde{v}_k+1)^{2L}\nonumber\\
  &&=\tilde{v}_k^{2L}\bar{a}(\tilde{v}_k)
  \frac{Q^{(1)}(\tilde{v}_k+1)}{Q^{(1)}(\tilde{v}_k)}, \quad k=1,\cdots, N, \label{bae_prime_01}\\
  &&\bar{a}(\tilde{\lambda}_l)Q(\tilde{\lambda}_l-1)
  Q^{(1)}(\tilde{\lambda}_l+1)+
  \bar{d}(\tilde{\lambda}_l)Q(\tilde{\lambda}_l)Q^{(1)}(\tilde{\lambda}_l-1)\nonumber\\
  &&=2h\tilde{\lambda}_l(\tilde{\lambda}_l-\frac{1}{2})
  Q(\tilde{\lambda}_l)Q(\tilde{\lambda}_l-1), \quad
  l=1,\cdots,N. \label{bae_prime_02}
\end{eqnarray}
According to Eq.\eqref{boundary-matrix3}, the energy spectrum of the Hamiltonian \eqref{Hamiltonian} can be expressed by
the eigenvalue $\Lambda(u)$ as
\begin{eqnarray}
\label{energy_prime_01}
  E &=&-\frac{1}{2}\frac{d\ln\Lambda(u)}{du}\Big|_{u=0}+\frac{1}{2\xi}
  -\frac{1-2\xi'}{2(1-\xi')}-2N+L-1\nonumber\\
  &=&-\sum_{k=1}^N\frac{1}{\tilde{v}_k(\tilde{v}_k+1)}-2N,
\end{eqnarray}
where the Bethe roots $\{\tilde{v}_k\}$ should satisfy the inhomogeneous BAEs \eqref{bae_prime_01}-\eqref{bae_prime_02}.

\section{Finite-size scaling analysis}
\label{sec3}

If the boundary parameters satisfy the constraint $h=0$, the inhomogeneous
$T-Q$ relation \eqref{eigenvalue} reduces to
\begin{eqnarray}
\label{new_tq}
  &&\Lambda_{hom}(u) = \omega_3(u)(\xi+u)(u+1)^{2L}
  \frac{Q(u-1)}{Q(u)}-u^{2L}\bar{a}(u)\nonumber\\
  &&\;\;\;\times\frac{Q(u-1)Q^{(1)}(u+1)}{Q(u)Q^{(1)}(u)}
  -u^{2L}\bar{d}(u)\frac{Q^{(1)}(u-1)}{Q^{(1)}(u)}.
\end{eqnarray}
Put $\tilde{v}_k=i\mu_k-\frac{1}{2}$ and $\tilde{\lambda}_l= i\lambda_l$. Then the Bethe roots $\{\mu_k\}$ and $\{\lambda_l\}$ should
satisfy the reduced Bethe ansatz equations
\begin{eqnarray}
  &&\frac{\mu_k-(\frac{1}{2}-\xi')i} {\mu_k+(\frac{1}{2}-\xi')i}
  \left(\frac{\mu_k-\frac{i}{2}}{\mu_k+\frac{i}{2}}\right)^{2L}\nonumber\\
   &&=-\prod_{j=1}^{M}\frac{\mu_k-\lambda_j-\frac{i}{2}}
   {\mu_k-\lambda_j+\frac{i}{2}}\frac{\mu_k+\lambda_j-\frac{i}{2}}
   {\mu_k+\lambda_j+\frac{i}{2}}, \quad k=1, \cdots, N, \label{BAE01}\\
  &&\frac{\lambda_l+\frac{i}{2}}{\lambda_l-\frac{i}{2}}
  \frac{\lambda_l-\xi'i}{\lambda_l+\xi'i}
  \frac{\lambda_l-\xi i}{\lambda_l+\xi i}
  =
  -\prod_{j=1}^N\frac{\lambda_l-\mu_j-\frac{i}{2}}{\lambda_l-\mu_j+\frac{i}{2}}
  \frac{\lambda_l+\mu_j-\frac{i}{2}}{\lambda_l+\mu_j+\frac{i}{2}}\nonumber\\
  &&\times\prod_{m=1}^{M}
  \frac{\lambda_l-\lambda_m+i}{\lambda_l-\lambda_m-i}
  \frac{\lambda_l+\lambda_m+i}{\lambda_l+\lambda_m-i}, \quad l=1, \cdots, M,\label{BAE02}
\end{eqnarray}
where $M$ is the number of Bethe roots $\{\lambda_l\}$.
We shall note that when the $T-Q$ relation is homogeneous, the integer $M$ could be smaller than $N$
because $N-M$ Bethe roots in the set of $\{\lambda_l\}$ tend to infinity and the associated constraints vanished in the BAEs \eqref{BAE01}-\eqref{BAE02} \cite{wang2015}.
According to Eq.\eqref{energy_prime_01}, we define the reduced energy $E_{hom}$ as
\begin{eqnarray}
\label{homo_eigenvalue}
  E_{hom} &=&-\frac{1}{2}\frac{d\ln\Lambda_{hom}(u)}{du}\Big|_{u=0}+\frac{1}{2\xi}
  -\frac{1-2\xi'}{2(1-\xi')}\nonumber\\
  &&-2N+L-1=\sum_{k=1}^N\frac{1}{\mu_k^2+\frac{1}{4}}-2N.
\end{eqnarray}
It is remarked that $E_{hom}$ is not the eigenvalue $E$ of Hamiltonian \eqref{Hamiltonian}.
In order to characterize the difference between $E_{hom}$ and $E$, we define a quantity
\begin{equation}
  \delta_e=|E-E_{hom}|,\label{homo_1eigenvalue}
\end{equation}
which measures the contribution of the inhomogeneous term.
The values of $\delta_e$ can be calculated as follows.
For the given system size $L$ and the filling factor $n=N/L$, we solve the BAEs (\ref{bae_prime_01})-(\ref{bae_prime_02}).
Substituting the solutions of Bethe roots into Eq.\eqref{energy_prime_01}, we obtain the values of $E$.
Similar, from the solutions of BAEs (\ref{BAE01})-(\ref{BAE02}) and the energy expression \eqref{homo_eigenvalue},
we obtain the values of $E_{hom}$. Substituting $E$ and $E_{hom}$ into \eqref{homo_1eigenvalue}, we arrive at the values of $\delta_e$.

Here, we focus on the ground state. In order to obtain the patterns of Bethe roots, we first consider the critical behavior of boundary fields.
From Eq.\eqref{boundary_parameters}, we know that the poles of the boundary parameters are $\xi=0$ and $\xi'=1$.
Thus we divide the boundary parameters into four different regimes: (i) $\xi>0$, $\xi'<1$, (ii) $\xi>0$, $\xi'>1$, (iii) $\xi<0$, $\xi'<1$ and (iv) $\xi<0$, $\xi'>1$.
In these regimes, the patterns of Bethe roots are different and we should consider them separately.
The second thing we mentioned is that the ground state energy and reduced one with finite system size $L$ can also be obtained by using the DMRG.
Then it is not necessary to solve the BAEs due to the complicated structure of Bethe roots.

\begin{figure}[htbp]
    \includegraphics[scale=0.68]{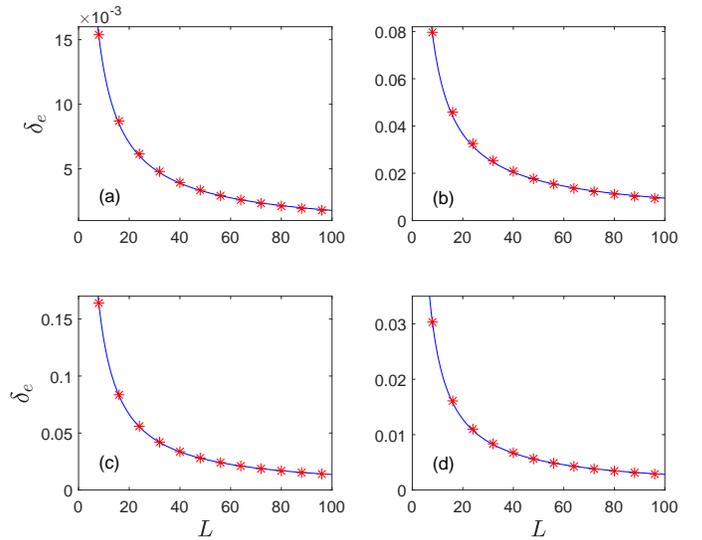}
  \caption{The values of $\delta_e$ versus the system size $L$. Here the filling factors $n=N/L=1$ and $\varphi=\theta'=\varphi'=0$. The constraint $h\neq0$ is achieved by putting $\theta=\pi/3$
  and $h=0$ by $\theta=0$. The data can be fitted as $\delta_e=\gamma L^\beta$, where (a) $\xi=0.413, \xi'=-3$, $\gamma=0.09045$ and $\beta=-0.8499$. (b) $\xi=0.413$, $\xi'=2.413$, $\gamma=0.4604$ and $\beta=-0.8403$. (c) $\xi=-0.413$, $\xi'=0.613$, $\gamma=1.271$ and $\beta=-0.9842$. (d) $\xi=0.4130$, $\xi'=2.413$, $\gamma=0.214$ and $\beta=-0.9379$. Due to the fact of $\beta<0$, when $L$ tends to infinity, the contribution of the inhomogeneous term tends to zero. }
  \label{figure01}
\end{figure}

In the above four regimes, we randomly choose the values of boundary parameters $\xi$ and $\xi'$.
In every regime, we choose two sets of boundary parameters $\{\theta, \varphi, \theta', \varphi'\}$.
One set satisfies the constraint $h=0$ while the other set does not.
Then we calculate the quantity $\delta_e$ by the DMRG.
The values of $\delta_e$ versus the system size $L$ in different regimes are shown in Fig.\ref{figure01}.
From the fitting, we find that $\delta_e$ and $L$ satisfy the power law, i.e., $\delta_e=\gamma L^\beta$.
Due to the fact of $\beta<0$, we conclude that
$\delta_e$ tends to zero when the system size $L$ tends to infinity, which means that the inhomogeneous term in the $T-Q$ relation \eqref{eigenvalue} can be neglected at the ground state in the thermodynamic limit.

\section{Surface energy}
\label{sec3.0}

In this section, we calculate the ground state energy and surface energy in the thermodynamic limit.
Because the patterns of Bethe roots depend on the boundary fields which have been divided into four regimes. We consider them separately.

\subsection{Regime (i): $\xi>0$ and $\xi'<1$}

From the analysis of BAEs \eqref{BAE01} and \eqref{BAE02}, we know that at the ground state,
there are $N$ Bethe roots $\{\mu_k\}$ and $N/2$ Bethe roots $\{\lambda_l\}$. Thus $M=N/2$. Meanwhile,
$\{\mu_k\}$ and $\{\lambda_l\}$ form the two-strings, i.e., the complex conjugate pairs
\begin{equation}\label{two-string}
  \mu_k=\lambda_l\pm\frac{i}{2}+\mathrm{o}(e^{-\varepsilon L}),\;\;\;
  l=1,\cdots,M, \;\;\; M=N/2,
\end{equation}
where $\lambda_l$ is position of two-string in the real axis. Without losing generality, $N$ is set to be even. $\mathrm{o}(e^{-\varepsilon L})$ means the finite size correction which can be neglected
if $L\rightarrow \infty$. From Eq.\eqref{two-string}, we know that the ground state of the system consists of the bounded singlet pairs with arbitrary spatial separation \cite{Bares-1991}.

Substituting $\mu_{k_1}=\lambda_l+\frac{i}{2}$, $\mu_{k_2}=\lambda_l-\frac{i}{2}$ into reduced BAEs \eqref{BAE01} and multiplying these two equations, we obtain
\begin{eqnarray}
  &&\frac{\lambda_l+i(\xi'-1)} {\lambda_l-i(\xi'-1)}
  \frac{\lambda_l+i\xi'} {\lambda_l-i\xi'}
  \left(\frac{\lambda_l-i}{\lambda_l+i}\right)^{2L}\nonumber\\
   &&=
  -\prod_{j=1}^{M}\frac{\lambda_l-\lambda_j-i}
   {\lambda_l-\lambda_j+i}\frac{\lambda_l+\lambda_j-i}
   {\lambda_l+\lambda_j+i},\quad l=1, \cdots, M.   \label{BAE01_new}
\end{eqnarray}
Rewrite BAEs \eqref{BAE02} as
\begin{eqnarray}
  && \prod_{j=1}^{M}\frac{\lambda_l-\lambda_j-i}
   {\lambda_l-\lambda_j+i}
   \frac{\lambda_l+\lambda_j-i}
   {\lambda_l+\lambda_j+i}=-
  \frac{\lambda_l-\frac{i}{2}}{\lambda_l+\frac{i}{2}}
  \frac{\lambda_l+\xi'i}{\lambda_l-\xi'i}\frac{\lambda_l+\xi i}{\lambda_l-\xi i}
  \nonumber\\
  &&\times\prod_{k=1}^N\frac{\lambda_l-\mu_k-\frac{i}{2}}
  {\lambda_l-\mu_k+\frac{i}{2}}
  \frac{\lambda_l+\mu_k-\frac{i}{2}}{\lambda_l+\mu_k+\frac{i}{2}}, \quad l=1, \cdots, M. \label{BAE02_new}
\end{eqnarray}
Substituting Eq.\eqref{BAE02_new} into \eqref{BAE01_new}, the Bethe ansatz equations become
\begin{eqnarray}
\label{BAE-string01}
  &&\left(\frac{\lambda_l-i}{\lambda_l+i}\right)^{2L} =
  -\frac{\lambda_l-\frac{i}{2}}{\lambda_l+\frac{i}{2}}
  \frac{\lambda_l+i(1-\xi')}{\lambda_l-i(1-\xi')}
  \frac{\lambda_l+i\xi}{\lambda_l-i\xi}\nonumber\\
  &&\times\prod_{j=1}^{M}\frac{\lambda_l-\lambda_j-i}{\lambda_l-\lambda_j+i}
  \frac{\lambda_l+\lambda_j-i}{\lambda_l+\lambda_j+i}, \quad l=1, \cdots, M.
\end{eqnarray}
Taking the logarithm of Eq.\eqref{BAE-string01}, we have
\begin{eqnarray}
\label{density-string01}
  &&2L\theta_2(\lambda_l)=2\pi I_l+\theta_1(\lambda_l)-\theta_{2(1-\xi')}(\lambda_l)-\theta_{2\xi}(\lambda_l)
  \nonumber\\
  && +\sum_{j=1}^{M}\theta_2(\lambda_l-\lambda_j)+
  \theta_2(\lambda_l+\lambda_j), \;\; l=1, \cdots, M,
\end{eqnarray}
where $\{I_l\}$ are the quantum numbers characterizing the ground state and take the values
\begin{eqnarray}
  &&\{I_l\}=\{-I_{max}, -I_{max}+1, \cdots, -I_{max}+(M-1), \nonumber\\
  && \quad I_{max}-(M-1), I_{max}-(M-1)+1, \cdots, I_{max}\},\nonumber\\
  && \quad I_{max}=L-M, \quad I_j\neq 0,
\end{eqnarray}
and the function $\theta_m(\lambda)$ is given by
\begin{eqnarray}
\theta_m(\lambda) = 2\arctan(2\lambda/m).
\end{eqnarray}
At half-filling $n=N/L=1$, the set of quantum numbers are $\{-\frac N2, -\frac N 2 +1, \cdots, -1, 1, \cdots, \frac N 2\}$.
If the filling factor $n<1$, the number of available occupation states is larger than the number of states of electrons.
Thus the lowest energy can be achieved by choosing $I_l$ as large as possible.

In the thermodynamic limit, i.e., $L\rightarrow\infty$ and $N/L$ is finite, the distribution of Bethe roots on the real axis tends to continuous.
Taking the derivative of Eq.\eqref{density-string01}, we obtain the density $\rho_1(\lambda)$ of Bethe roots as
\begin{eqnarray}
\label{density-02}
\rho_1(\lambda) &=& a_2(\lambda)-\frac{1}{2L}\left[a_1(\lambda)-a_{2(1-\xi')}(\lambda)-
  a_{2\xi}(\lambda)\right]\nonumber\\
&&-\Big[\int_{-\infty}^{-Q_0} +\int_{Q_0}^{\infty}\Big]a_2(\lambda-\mu)\rho_1(\mu)d\mu,
\end{eqnarray}
where
\begin{eqnarray}
a_m(\lambda)=\frac{1}{2\pi}\frac{m}{\lambda^2+m^2/4},
\end{eqnarray}
and $Q_0$ is determined by the constraint
\begin{equation}\label{n1_sun}
  \left[\int_{-\infty}^{-Q_0}+\int_{Q_0}^\infty\right]
  \rho_1(\lambda)d\lambda=\frac{N}{2L}=\frac{n}{2}.
\end{equation}
Taking the Fourier transformation of Eq.\eqref{density-02}, we have
\begin{eqnarray}
  \tilde{\rho}_1(w) &=& \frac{e^{-|w|}}{1+e^{-|w|}}-
  \frac{1}{2L}\frac{e^{-\frac{|w|}{2}}-e^{-(1-\xi')|w|}-e^{-\xi|w|}}
  {1+e^{-|w|}}\nonumber\\
  &&+\int_{-Q_0}^{Q_0}\frac{e^{-|w|}e^{-iw(\lambda-\mu)}}{1+e^{-|w|}}
  \rho_1(\mu)d\mu.
\end{eqnarray}
Based on it, we obtain the ground state energy as
\begin{equation}\label{energy-general-01}
  E_1=-2L\left[\int_{-\infty}^{-Q_0}+\int_{Q_0}^{\infty}\right]
  d\lambda\rho_{1}(\lambda)\left[2-\frac{1}{1+\lambda^2}\right].
\end{equation}

\begin{figure}[htbp]
  \centering
    \includegraphics[scale=0.85]{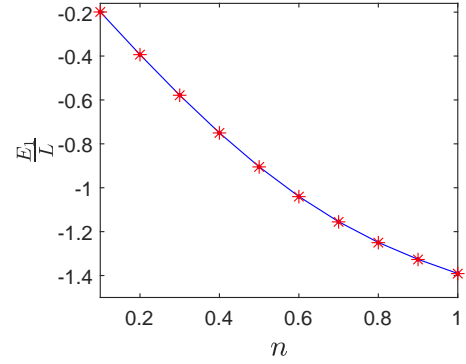}
  \caption{The ground state energy density $E_1/L$ versus the filling factor $n$. The curve is the result calculated from Eq.\eqref{energy-general-01},
  and the red stars are the results by using the DMRG and the finite size scaling analysis. Here, the model parameters are chosen as $\xi=1.9$, $\xi'=0.5$, $\theta=\pi/3$ and $\phi=\theta'=\phi'=0$.}
  \label{figure-anyn-01}
\end{figure}

For the given boundary parameters, we solve Eq.\eqref{density-02} with the constraint \eqref{n1_sun} numerically.
Substituting the solution into Eq.\eqref{energy-general-01}, we obtain the value of $E_1$. The ground state energy density $E_1/L$ versus the filling factor $n$ is shown in Fig.\ref{figure-anyn-01} as the blue curve.
In order to check the correction of analytical expression \eqref{energy-general-01}, we also diagonalize the Hamiltonian \eqref{Hamiltonian} with fixed filling factors by the DMRG.
From the finite size scaling analysis of data, we also obtain the values of $E_1/L$ and the results are shown in Fig.\ref{figure-anyn-01} as the red stars.
From Fig.\ref{figure-anyn-01}, we see that the analytical and DMRG results agree with each other very well.

At the half-filling, $Q_0=0$ and Eq.\eqref{density-02} reduces to
\begin{eqnarray}
  \rho_{01}(\lambda) &=& a_2(\lambda)-\frac{1}{2L}\Big[a_1(\lambda)-a_{2(1-\xi')}(\lambda)-
  a_{2\xi}(\lambda)\nonumber\\
  &&+\delta(\lambda)\Big]-\int_{-\infty}^\infty a_2(\lambda-\mu)\rho_{01}(\mu)d\mu. \label{20211102-1}
\end{eqnarray}
We shall note that the introducing of function $\delta(\lambda)$ is due to the existence of hole in the set of quantum numbers.
From the quantum number of hole, we can obtain a set of solutions of Bethe roots but the corresponding wave function is zero \cite{wang2015}.
The Fourier transformation of Eq.\eqref{20211102-1} gives
\begin{eqnarray*}
  \tilde{\rho}_{01}(w) &=& \frac{e^{-|w|}}{1+e^{-|w|}}-
  \frac{e^{-\frac{|w|}{2}}-e^{-(1-\xi')|w|}-e^{-\xi|w|}+1}
  {2L(1+e^{-|w|})}.
\end{eqnarray*}
\begin{figure}[htbp]
  \centering
    \includegraphics[scale=0.67]{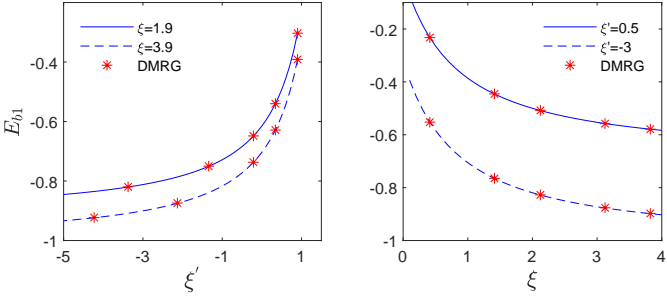}
  \caption{The surface energy $E_{b1}$ at the half-filling. The blue curves are the results calculated from analytical expression \eqref{surface-energy-01}, and the red stars are the ones obtained by the finite size scaling analysis ($L\rightarrow \infty$) of DMRG data. Here, the boundary parameters are chosen as $\theta=\pi/3$, $\phi=\theta'=\phi'=0$ and $n=1$.}
  \label{figure-surface-01}
\end{figure}
Accordingly, we obtain the ground state energy as
\begin{eqnarray}
&&  E_{01} =E_p+E_{b1},  \\
&& E_p= -2N+L\int_{-\infty}^\infty\frac{e^{-2|w|}}{1+e^{-|w|}}dw=-2L\ln 2,  \\
&& E_{b1}= -\int_{-\infty}^\infty\frac{e^{-|w|}}{2(1+e^{-|w|})} \nonumber \\
&& \qquad \quad \times [e^{-\frac{|w|}{2}}-e^{-\xi|w|}-e^{-(1-\xi')|w|}+1] dw,\label{surface-energy-01}
\end{eqnarray}
where $E_p$ is the ground state energy of one-dimensional supersymmetric $t-J$ model with periodic boundary conditions \cite{Bares-1991}
and $E_{b1}$ is the surface energy of the Hamiltonian \eqref{Hamiltonian}.
The results are shown in Fig.\ref{figure-surface-01}, where the blue curves are the surface energies of the system with different boundary parameters calculated by the analytical expression \eqref{surface-energy-01}
and the red stars are the results obtained by the finite size scaling analysis ($L\rightarrow \infty$) of DMRG data. From Fig.\ref{figure-surface-01}, we see that the analytical results agree with the numerical ones very well.

\subsection{Regime (ii): $\xi>0$ and $\xi'>1$}

In the regime of $\xi>0$ and $\xi'>1$, detailed analysis of BAEs \eqref{BAE01} and \eqref{BAE02} gives that at the ground state,
there are $N$ Bethe roots $\{\mu_k\}$ and $N/2-1$ Bethe roots $\{\lambda_l\}$. In the thermodynamic limit,
the pattern of Bethe roots $\{\mu_k\}$ includes one real root $\mu_{N-1}$, one pure imaginary root $\mu_{N}=i(\xi'-\frac{1}{2})$ and
$N/2-1$ two-strings, which correspond the bound states,
\begin{eqnarray}
\mu_{k} = v_l \pm\frac{i}{2},\quad l=1, \cdots, \frac N 2 -1, \label{20211103-1}
\end{eqnarray}
where $v_l$ is position of two-string in the real axis. Meanwhile, all the Bethe roots $\{\lambda_l\}$ are real. We
shall note that different from the pattern of Bethe roots in the regime (i), the
present positions $\{v_l\}$ of two-strings do not equal to the Bethe roots $\{\lambda_l\}$.

Substituting $\mu_{k_1}=v_l+\frac{i}{2}$, $\mu_{k_2}=v_l-\frac{i}{2}$ into reduced BAEs \eqref{BAE01} and multiplying these two equations, we obtain
\begin{eqnarray}
&&\left(\frac{v_l-i}{v_l+i}\right)^{2L}
\frac{v_l+i\xi'}{v_l-i\xi'}
\frac{v_l+i(\xi'-1)}{v_l-i(\xi'-1)}=
\prod_{j=1}^{M_2}\frac{v_l-\lambda_j-i}{v_l-\lambda_j+i}
\nonumber\\
&& \times\frac{v_l+\lambda_j-i}{v_l+\lambda_j+i},
\quad l=1,\cdots,M_2,  \quad M_2=\frac N 2 -1. \label{20211105-1}
\end{eqnarray}
Substituting the real Bethe root $\mu_{N-1}$ into BAEs \eqref{BAE01}, we obtain
\begin{eqnarray}
  &&\left(\frac{\mu_{N-1}-\frac{i}{2}}{\mu_{N-1}+\frac{i}{2}}\right)^{2L}
  \frac{\mu_{N-1}-i(\frac{1}{2}-\xi')}{\mu_{N-1}+i(\frac{1}{2}-\xi')}\nonumber\\
&&=
  -\prod_{j=1}^{M_2}\frac{\mu_{N-1}-\lambda_j-\frac{i}{2}}
  {\mu_{N-1}-\lambda_j+\frac{i}{2}}
\frac{\mu_{N-1}+\lambda_j-\frac{i}{2}}{\mu_{N-1}+\lambda_j+\frac{i}{2}}.\label{20211105-2}
\end{eqnarray}
It is easy to check that $\mu_N$ satisfy the BAEs \eqref{BAE01} automatically.
The solution $\mu_N$ is the boundary string because it is determined by the boundary parameter $\xi'$.
Substituting the patterns of Bethe roots $\{\mu_k\}$ and $\{\lambda_l\}$ into BAEs \eqref{BAE02}, we have
\begin{eqnarray}
&&\frac{\lambda_l+\frac{i}{2}}{\lambda_l-\frac{i}{2}}
\frac{\lambda_l-i\xi}{\lambda_l+i\xi}
=-\frac{\lambda_l+i(\xi'-1)}{\lambda_l-i(\xi'-1)}
\prod_{j=1}^{M_2}\frac{\lambda_l-v_j-i}{\lambda_l-v_j+i}
\nonumber\\
&&\times\frac{\lambda_l+v_j-i}{\lambda_l+v_j+i}
\frac{\lambda_l-\mu_{N-1}-\frac{i}{2}}{\lambda_l-\mu_{N-1}+\frac{i}{2}}
\frac{\lambda_l+\mu_{N-1}-\frac{i}{2}}{\lambda_l+\mu_{N-1}+\frac{i}{2}}
\nonumber\\
&&\times\prod_{k=1}^{M_2}\frac{\lambda_l-\lambda_k+i}{\lambda_l-\lambda_k-i}
\frac{\lambda_l+\lambda_k+i}{\lambda_l+\lambda_k-i}, \quad l=1,\cdots,M_2.\label{20211105-3}
\end{eqnarray}
In the thermodynamic limit, taking the logarithm then the derivative of BAEs \eqref{20211105-1}-\eqref{20211105-3},
we find that the density of real parts of two-strings $\rho_{21}(v)$ and
the density of $\lambda$-root $\rho_{22}(\lambda)$ should satisfy following coupled integral equations
\begin{eqnarray}
  \label{BAEbou02-01}
  &&a_2(v)=\frac{1}{2L}
  \Big[a_{2\xi'}(v)+a_{2(\xi'-1)}(v)+\delta(v-\mu_{N-1})\nonumber\\
  &&\quad +\delta(v+\mu_{N-1})
  \Big]+\rho_{21}(v)\nonumber\\
  &&\quad+\left[\int_{-\infty}^{-B}+\int_B^{\infty}\right]
  a_2(v-\lambda)\rho_{22}(\lambda)d\lambda, \\
  &&\frac{1}{2L}\Big[a_1(\lambda)-a_{2\xi}(\lambda)-
  a_{2(\xi'-1)}(\lambda)+a_1(\lambda-\mu_{N-1})
  \nonumber\\
  &&\quad+a_1(\lambda+\mu_{N-1})\Big]+\left[\int_{-\infty}^{-Q_0}+\int_{Q_0}^{\infty}\right]
  a_2(\lambda-v)\rho_{21}(v)dv\nonumber\\
  &&= \frac{1}{2L}\left[\delta(\lambda-\mu_{N-1})
  +\delta(\lambda+\mu_{N-1})\right]\nonumber\\
  &&\quad+\left[\int_{-\infty}^{-B}+\int_{B}^{\infty}\right]
  a_2(\lambda-u)\rho_{22}(u)du,
  \label{BAEbou02-02}
\end{eqnarray}
where $\mu_{N-1}$ characterizes the position of hole in the sea of two-strings, the integration limit $Q_0$ is determined by the constraint
 \begin{equation}\label{n2_sun}
  \left[\int_{-\infty}^{-Q_0}+\int_{Q_0}^\infty\right]
  \rho_{21}(v)dv=\frac{N-2}{2L}=\frac{n}{2}-\frac{1}{L}.
\end{equation}
Substituting Eq.\eqref{BAEbou02-02} into \eqref{BAEbou02-01}, we arrive at the
integral equation for the density $\rho_{21}(v)$
\begin{eqnarray}
\label{density-02_2}
  &&\rho_{21}(v) = a_2(v)-\frac{1}{2L}
  [a_1(v)-a_{2\xi}(v)+a_{2\xi'}(v)
  \nonumber\\
  &&\qquad +a_1(v-\mu_{N-1})+a_1(v+\mu_{N-1})]\nonumber\\
  &&\qquad
  -\Big[\int_{-\infty}^{-Q_0}+\int_{Q_0}^\infty\Big]a_2(v-\mu)\rho_{21}(\mu)d\mu.
\end{eqnarray}
The Fourier transformation of Eq.\eqref{density-02_2} reads
\begin{eqnarray}
  && \tilde{\rho}_{21}(w) =
  \frac{e^{-\xi|w|}-e^{-\xi'|w|}-[1+2\cos(w\mu_{N-1})]e^{-\frac{|w|}{2}}}
  {2L(1+e^{-|w|})}\nonumber\\
  &&\quad +\frac{e^{-|w|}}{1+e^{-|w|}}+
  \int_{-Q_0}^{Q_0}\frac{e^{-|w|}e^{-iw(\lambda-\mu)}}{1+e^{-|w|}}
  \rho_{21}(\mu)d\mu.\label{energy-g2eneral-02}
\end{eqnarray}
Using Eq.\eqref{energy-g2eneral-02}, we obtain the ground state energy as
\begin{eqnarray}\label{energy-general-02}
  E_2&=&-2N+2L\pi\left[\int_{-\infty}^{-Q_0}+\int_{Q_0}^\infty\right]a_2(v)
  \rho_{21}(v)dv\nonumber\\
  &&+\frac{1}{\mu_{N-1}^2+\frac{1}{4}}+
  \frac{1}{\xi'-\xi'^2}.
\end{eqnarray}
We see that with the increasing of $\mu_{N-1}$, the ground state energy is decreasing. Thus, $\mu_{N-1}$
should be put at the infinity, i.e., $\mu_{N-1}\rightarrow\infty$, to minimize the energy and obtain the stable ground state.

\begin{figure}[htbp]
  \centering
    \includegraphics[scale=0.85]{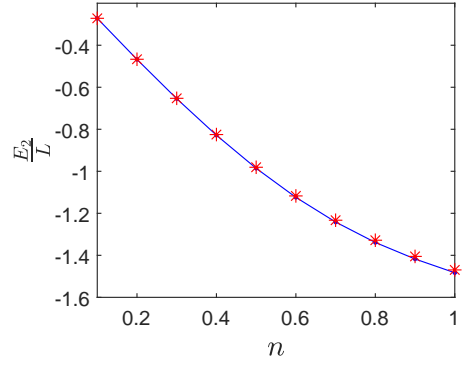}
  \caption{The ground state energy density $E_2/L$ versus the filling factor $n$. The curve is obtained from Eq.\eqref{energy-general-02} and the red stars are
  obtained by the finite size scaling analysis of DMRG data. Here, the model parameters are chosen as $\xi=0.9$, $\xi'=1.123$, $\theta=\pi/3$ and $\phi=\theta'=\phi'=0$.}
  \label{figure-anyn-02}
\end{figure}

Now, we check the correction of above results. For the given boundary parameters in this regime, we solve Eq.\eqref{density-02_2} numerically, where the integration limit $Q_0$ thus
the filling factor $n$ satisfies the constraint \eqref{n2_sun},
and obtain the value of the density $\rho_{21}(v)$. Substituting $\rho_{21}(v)$ into Eq.\eqref{energy-general-02}, we obtain the ground state energy $E_2$.
The energy per site $E/L$ versus the filling factor $n$ is shown in Fig.\ref{figure-anyn-02} as the blue curve.
On the other hand, by using the DMRG, we diagonalize the Hamiltonian \eqref{Hamiltonian} with same boundary parameters.
From the finite size scaling analysis of DMRG data, we also obtain the ground state energy density $E/L$ which are shown in Fig.\ref{figure-anyn-02} as the red stars.
From Fig.\ref{figure-anyn-02}, we see that the analytical and DMRG results agree with each other very well.

\begin{figure}[htbp]
  \centering
    \includegraphics[scale=0.67]{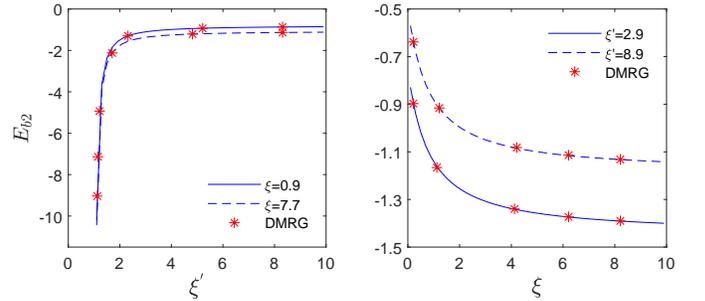}
  \caption{ The surface energies of the system with different boundary parameters. The curves are the results calculated from Eq.\eqref{surface-02}, and the red stars are obtained by the DMRG and
  finite size scaling analysis. Here, the model parameters are chosen as $\theta=\pi/3$, $\phi=\theta'=\phi'=0$ and $n=1$.}
  \label{figure-surface-02}
\end{figure}

At the half-filling, Eq.\eqref{energy-g2eneral-02} reduces to
\begin{eqnarray}
&&  \tilde{\rho}_{02}(w) =[2L(1+e^{-|w|})]^{-1}\{2Le^{-|w|}-e^{-\xi'|w|} \nonumber \\
&&\quad   +e^{-\xi|w|}-[1+2\cos(w\mu_{N-1})]e^{-\frac{|w|}{2}}+1\}.
\end{eqnarray}
Considering the fact that $\mu_{N-1}=\infty$ at the ground state, we obtain the surface energy as
\begin{eqnarray}
  E_{b2} &=&-\int_{-\infty}^\infty\frac{e^{-|w|}}{2}
  \frac{e^{-\frac{|w|}{2}}-e^{-\xi|w|}+e^{-\xi'|w|}+1}{1+e^{-|w|}}dw
  \nonumber\\
  &&+\frac{1}{\xi'-\xi'^2}.
  \label{surface-02}
\end{eqnarray}
The results are shown in Fig.\ref{figure-surface-02}, where
the curves are the surface energies of the system with different boundary parameters calculated from the analytical relation \eqref{surface-02}
and the red stars are the ones obtained by the DMRG and finite size scaling analysis. We see that the analytical results agree with the numerical ones very well.

\subsection{Regime (iii): $\xi<0$ and $\xi'<1$}

If the boundary parameters belong to the regime of $\xi<0$ and $\xi'<1$, the BAEs \eqref{BAE01} and \eqref{BAE02} gives that there are $N$ Bethe roots $\{\mu_k\}$ and $N/2$ Bethe roots $\{\lambda_l\}$ at the ground state.
In the thermodynamic limit, the pattern of Bethe roots $\{\mu_k\}$ includes one real root $\mu_{N-1}$, one boundary string $\mu_{N}=i(\frac{1}{2}-\xi)$ and
$N/2-1$ two-strings with the form of
\begin{eqnarray}
\mu_{k} =v_j\pm\frac{i}{2},\quad j=1,\cdots,\frac N 2 -1,
\end{eqnarray}
where $v_j$ is the position of two-string in the real axis. Meanwhile, the pattern of Bethe roots $\{\lambda_l\}$ includes $N/2-1$ real
roots and one boundary string $\lambda_{N/2}=-i\xi$.

It is easy to check that the boundary strings $\mu_N$ and $\lambda_{N/2}$ satisfy the BAEs naturally.
Then the constraints of undetermined Bethe roots are
\begin{eqnarray}
&&\left(\frac{v_j-i}{v_j+i}\right)^{2L}
\frac{v_j+i\xi'}{v_j-i\xi'}
\frac{v_j+i(\xi'-1)}{v_j-i(\xi'-1)}=
\frac{v_j-i\xi-i}{v_j-i\xi+i}\nonumber\\
&&
\times\frac{v_j+i\xi-i}{v_j+i\xi+i}\prod_{l=1}^{M_2}\frac{v_j-\lambda_l-i}{v_j-\lambda_l+i}
\frac{v_j+\lambda_l-i}{v_j+\lambda_l+i},\nonumber\\
  && j=1,\cdots, M_2, \quad M_2=\frac N 2 -1, \label{BAEbou03-1}\\
  &&\left(\frac{\mu_{N-1}-\frac{i}{2}}{\mu_{N-1}+\frac{i}{2}}\right)^{2L}
  \frac{\mu_{N-1}-i(\frac{1}{2}-\xi')}{\mu_{N-1}+i(\frac{1}{2}-\xi')}\nonumber\\
&&=
-\prod_{l=1}^{M_2}\frac{\mu_{N-1}-\lambda_l-\frac{i}{2}}{\mu_{N-1}-\lambda_l+\frac{i}{2}}
\frac{\mu_{N-1}+\lambda_l-\frac{i}{2}}{\mu_{N-1}+\lambda_l+\frac{i}{2}},   \\
&&\frac{\lambda_j+\frac{i}{2}}{\lambda_j-\frac{i}{2}}
\frac{\lambda_j-i\xi'}{\lambda_j+i\xi'}
=-
\prod_{l=1}^{M_2}\frac{\lambda_j-v_l-i}{\lambda_j-v_l+i}
\frac{\lambda_j+v_l-i}{\lambda_j+v_l+i}
\nonumber\\
&&\times\prod_{k=1}^{M_2}\frac{\lambda_j-\lambda_k+i}{\lambda_j-\lambda_k-i}
\frac{\lambda_j+\lambda_k+i}{\lambda_j+\lambda_k-i}
\frac{\lambda_j+i(1+\xi)}{\lambda_j-i(1+\xi)}\nonumber\\
&&\times\frac{\lambda_j-\mu_{N-1}-\frac{i}{2}}{\lambda_j-\mu_{N-1}+\frac{i}{2}}
\frac{\lambda_j+\mu_{N-1}-\frac{i}{2}}{\lambda_j+\mu_{N-1}+\frac{i}{2}},
j=1,\cdots, M_2.
\label{BAEbou03}
\end{eqnarray}
In the thermodynamic limit, taking the logarithm then the derivative of BAEs \eqref{BAEbou03-1}-\eqref{BAEbou03}, we obtain that the densities of Bethe roots should
satisfy the coupled integral equations
\begin{eqnarray}
&&  a_2(v) = \frac{1}{2L}
  \left[a_{2\xi'}(v)-a_{2(1-\xi')}(v)+\delta(v-\mu_{N-1})
  \right.\nonumber\\
&& \;\; \left.+\delta(v+\mu_{N-1})+a_{2(\xi+1)}(v)+a_{2(1-\xi)}(v)
  \right]+\rho_{31}(v)\nonumber\\
&&\;\; +\left[\int_{-\infty}^{-B}+\int_B^{\infty}\right]
  a_2(v-\lambda)\rho_{32}(\lambda)d\lambda,\label{BAEbou03-01}\\
&&\frac{1}{2L}\left[a_1(\lambda)-a_{2\xi'}(\lambda)-
  a_{2(\xi+1)}(\lambda)+a_1(\lambda-\mu_{N-1})\right.
  \nonumber\\
  &&
  \;\;\left.+a_1(\lambda+\mu_{N-1})\right]+\left[\int_{-\infty}^{-Q_0}+\int_{Q_0}^{\infty}\right]
  a_2(\lambda-v)\rho_{31}(v)dv\nonumber\\
  &&= \frac{1}{2L}\left[\delta(\lambda-\mu_{N-1})
  +\delta(\lambda+\mu_{N-1})\right]\nonumber\\
  &&\;\;+
  \left[\int_{-\infty}^{-B}+\int_{B}^{\infty}\right]
  a_2(\lambda-u)\rho_{32}(u)du,
  \label{BAEbou03-02}
\end{eqnarray}
where $\rho_{31}(v)$ is the density of real parts of two-strings $\{v_j\}$, $\rho_{32}(\lambda)$ is the density of Bethe roots $\{\lambda_j\}$,
$\mu_{N-1}$ characterizes the hole in the sea of two-strings, the integration limit $Q_0$ is determined by the constraint
 \begin{equation}\label{n3_sun}
  \left[\int_{-\infty}^{-Q_0}+\int_{Q_0}^\infty\right]
  \rho_{31}(v)dv=\frac{N-2}{2L}=\frac{n}{2}-\frac{1}{L}.
\end{equation}
Substituting Eq.\eqref{BAEbou03-02} into \eqref{BAEbou03-01}, we obtain
\begin{eqnarray}
\label{density-03_2}
  &&\rho_{31}(v) = a_2(v)-\frac{1}{2L}
  \left[a_1(v)-a_{2(1-\xi')}(v)+a_{2(1-\xi)}(v)\right]
  \nonumber\\
  &&\qquad -\frac{1}{2L}\left[a_1(v-\mu_{N-1})+a_1(v+\mu_{N-1})\right]\nonumber\\
  &&\qquad -\Big[\int_{-\infty}^\infty -
  \int_{-Q_0}^{Q_0}\Big]a_2(v-\mu)\rho_{31}(\mu)d\mu.
\end{eqnarray}
The Fourier transformation of Eq.\eqref{density-03_2} gives
\begin{eqnarray}
  &&\tilde{\rho}_{31}(w) = \frac{e^{-|w|}}{1+e^{-|w|}}+ \int_{-Q_0}^{Q_0}\frac{e^{-|w|}e^{-iw(\lambda-\mu)}}{1+e^{-|w|}}
  \rho_{31}(\mu)d\mu\nonumber\\
  && -
  \frac{[1+2\cos(w\mu_{N-1})]e^{-\frac{|w|}{2}}-e^{-(1-\xi')|w|}+e^{-(1-\xi)|w|}}
  {2L(1+e^{-|w|})}. \nonumber
\end{eqnarray}
From above equation, we obtain the ground state energy
\begin{eqnarray}\label{energy-general-03}
  E_3&=&-2N+2L\pi\left[\int_{-\infty}^{-Q_0}+\int_{Q_0}^\infty\right]a_2(v)
  \rho_{31}(v)d v\nonumber\\
  &&+\frac{1}{\mu_{N-1}^2+\frac{1}{4}}+
  \frac{1}{\xi-\xi^2}.
\end{eqnarray}
Again, $\mu_{N-1}$ should tend to infinity to touch the stable ground state.

\begin{figure}[htbp]
  \centering
    \includegraphics[scale=0.85]{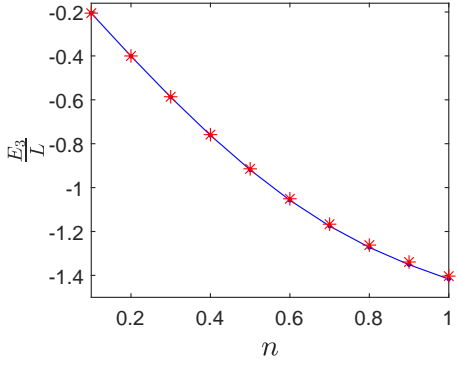}
  \caption{The ground state energy density $E_3/L$ versus the filling factor $n$.
  The curve is the result calculated from analytical expression \eqref{energy-general-03} and the red stars are obtained by the DMRG and finite size scaling.
  Here, the model parameters are chosen as $\xi=-0.9$, $\xi'=-0.9$, $\theta=\pi/3$ and $\phi=\theta'=\phi'=0$.}
  \label{figure-anyn-03}
\end{figure}

Now, we check the correction of analytical expression \eqref{energy-general-03}. For the given boundary parameters in regime (iii), we
solve the integral equation \eqref{density-03_2} numerically, where the integration limit $Q_0$ thus the filling factor $n$ satisfies the constraint \eqref{n3_sun}, and
obtain the value of density $\rho_{31}(v)$. Substituting $\rho_{31}(v)$ into \eqref{energy-general-03}, we obtain the ground state energy $E_3$.
The energy per site $E_3/L$ versus the filling factor $n$ is shown in Fig.\ref{figure-anyn-03} as the blue curve.
On the other hand, by using the DMRG, we diagonalizing the Hamiltonian \eqref{Hamiltonian} with same boundary parameters. From
the finite size scaling analysis of DMRG data, we also obtain the ground state energy density $E_3/L$ which are shown in Fig.\ref{figure-anyn-03} as the red stars.
From Fig.\ref{figure-anyn-03}, we see that all the results are consistent with each other very well.

\begin{figure}[htbp]
  \centering
    \includegraphics[scale=0.67]{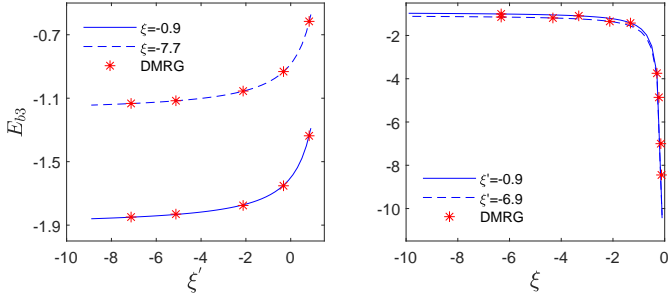}
  \caption{ The surface energies of the system in the thermodynamic limit. The curves are the results calculated from Eq.\eqref{surface-03} and the red stars are
  the ones obtained by the DMRG. Here, the boundary parameters are chosen as $\theta=\pi/3$, $\phi=\theta'=\phi'=0$ and $n=1$.}
  \label{figure-surface-03}
\end{figure}

At the half filling, the density $\tilde{\rho}_{03}(w)$ satisfies
\begin{eqnarray}
 && \tilde{\rho}_{03}(w) = \frac{e^{-|w|}}{1+e^{-|w|}}
  -\frac{1}{L}\frac{\cos(w\mu_{N-1})e^{-\frac{|w|}{2}}}{1+e^{-|w|}}
  \nonumber\\
  &&\quad -
  \frac{1}{2L}\frac{e^{-\frac{|w|}{2}}-e^{-(1-\xi')|w|}+e^{-(1-\xi)|w|}+1}
  {1+e^{-|w|}}.
\end{eqnarray}
Considering the fact $\mu_{N-1}=\infty$ and using the similar procedure, we obtained the surface energy as
\begin{eqnarray}
\label{surface-03}
 && E_{b3} =\int_{-\infty}^\infty
  \frac{e^{(\xi'-2)|w|}-e^{-\frac{3|w|}{2}}-e^{(\xi-2)|w|}-e^{-|w|}}{2(1+e^{-|w|})}dw
  \nonumber\\
  &&\qquad\;\; +\frac{1}{\xi-\xi^2}.
\end{eqnarray}
The results are shown in Fig.\ref{figure-surface-03}, where the curves are the surface energies calculated by using the expression \eqref{surface-03} with the given boundary parameters in regime (iii),
and the red stars are data obtained by using the DMRG and finite size scaling analysis. From Fig.\ref{figure-surface-03}, we see that the analytical results and
numerical ones agree with each other very well.

\subsection{Regime (iv): $\xi<0$ and $\xi'>1$}

In the regime of $\xi<0$ and $\xi'>1$, from the analysis of BAEs \eqref{BAE01} and \eqref{BAE02}, we obtain that
there are $N$ Bethe roots $\{\mu_k\}$ and $N/2$ Bethe roots $\{\lambda_l\}$ at the ground state.
In the thermodynamic limit, the pattern of Bethe roots $\{\mu_j\}$ includes two boundary strings, $\mu_{N-1}=i(\xi'-\frac{1}{2})$ and $\mu_N=-i(\xi-\frac{1}{2})$,
while the pattern of $\{\lambda_l\}$ includes one boundary string, $\lambda_{N/2}=-i\xi$.
The rest $N-2$ $\{\mu_k\}$ and $\frac N 2-1$ $\{\lambda_l\}$ form the two-strings with the form of
\begin{eqnarray}
\mu_{k}=\lambda_j \pm\frac{i}{2},\quad j=1,\cdots,\frac N 2-1,
\end{eqnarray}
where all the $\{\lambda_j\}$ are real.

\begin{figure}[htbp]
  \centering
    \includegraphics[scale=0.85]{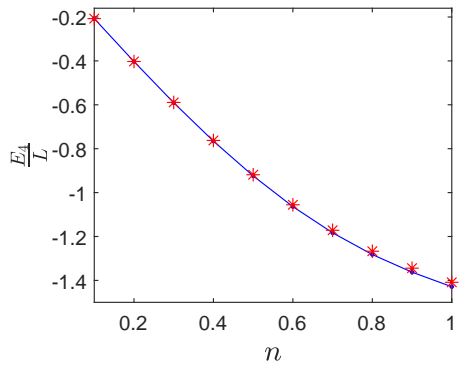}
  \caption{The ground state energy density $E_4/L$ versus the filling factor $n$. The curve is the result calculated from Eq.\eqref{energy-general-04} and
   the red stars are the data obtained by the DMRG. Here, $\xi=-0.9$, $\xi'=2.9$, $\theta=\pi/3$ and $\phi=\theta'=\phi'=0$.}
  \label{figure-anyn-04}
\end{figure}

Substitution these patterns into BAEs \eqref{BAE01}-\eqref{BAE02} and using the similar technique as in regime (i), we obtain
\begin{eqnarray}
  &&\hspace{-0.5cm}\left(\frac{\lambda_j-i}{\lambda_j+i}\right)^{2L} =-
  \frac{\lambda_j-\frac{i}{2}}{\lambda_j+\frac{i}{2}}
  \frac{\lambda_j-i(1-\xi)}{\lambda_j+i(1-\xi)}
  \frac{\lambda_j-i\xi'}{\lambda_j+i\xi'}\nonumber\\
  &&\hspace{-0.5cm}\times\prod_{k=1}^{M-1}\frac{\lambda_j-\lambda_k-i}{\lambda_j-\lambda_k+i}
  \frac{\lambda_j+\lambda_k-i}{\lambda_j+\lambda_k+i},  j=1,\cdots,M-1,\label{BAEbou04}
\end{eqnarray}
which gives that the density of Bethe roots on the real axis, $\rho_4(\lambda)$,
should satisfy the integral equation
\begin{eqnarray}
  \rho_4(\lambda) &=& a_2(\lambda)-\frac{1}{2L}\left[a_1(\lambda)+a_{2(1-\xi)}(\lambda)+
  a_{2\xi'}(\lambda)\right]\nonumber\\
  &&-\Big[\int_{-\infty}^{-Q_0}+\int_{Q_0}^\infty\Big]a_2(\lambda-\mu)\rho_4(\mu)d\mu,
  \label{bou04-density}
\end{eqnarray}
where $Q_0$ is determined by the constraint
\begin{equation}\label{n4_sun}
  \left[\int_{-\infty}^{-Q_0}+\int_{Q_0}^\infty\right]
  \rho_4(\lambda)d\lambda=\frac{N-2}{2L}=\frac{n}{2}-\frac{1}{L}.
\end{equation}
The Fourier transformation of Eq.\eqref{bou04-density} reads
\begin{eqnarray}
  \tilde{\rho}_4(w) &=& \frac{e^{-|w|}}{1+e^{-|w|}}-
  \frac{1}{2L}\frac{e^{-\frac{|w|}{2}}+e^{-(1-\xi)|w|}+e^{-\xi'|w|}}
  {1+e^{-|w|}}  \nonumber\\
  &&+\int_{-Q_0}^{Q_0}\frac{e^{-|w|}e^{-iw(\lambda-\mu)}}{1+e^{-|w|}}
  \rho_4(\mu)d\mu.
\end{eqnarray}
From it, we obtain the ground state energy
\begin{eqnarray}\label{energy-general-04}
  E_4&=&-2N+2L\pi\left[\int_{-\infty}^{-Q_0}+\int_{Q_0}^\infty\right]a_2(\lambda)
  \rho_4(\lambda)d\lambda\nonumber\\
  &+&\frac{1}{\xi'-\xi'^2}+
  \frac{1}{\xi-\xi^2}.
\end{eqnarray}
The ground state energy density $E_4/L$ versus the filling factor $n$ is shown in Fig.\ref{figure-anyn-04},
where the blue curve is the result obtained by the analytical expressions \eqref{bou04-density}, \eqref{n4_sun} and \eqref{energy-general-04},
and the red stars are the ones calculated by the DMRG. From Fig.\ref{figure-anyn-04}, we see that the analytical result \eqref{energy-general-04}
agrees with the DMRG result.

\begin{figure}[htbp]
  \centering
    \includegraphics[scale=0.67]{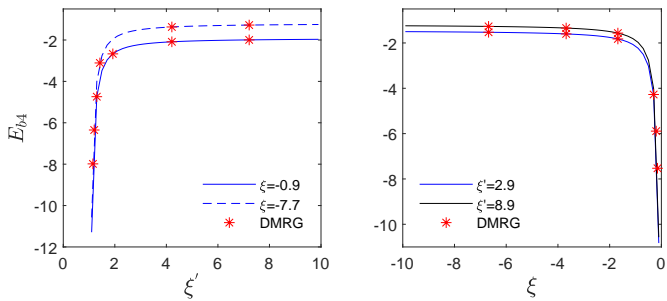}
  \caption{ The surface energies of the system with given boundary parameters in regime (iv). The curves are the results calculated from Eq.\eqref{surface-energy-bou04} and
  the red stars are the ones obtained by the DMRG and finite size scaling analysis. Here, $n=1$, $\theta=\pi/3$ and $\phi=\theta'=\phi'=0$.}
  \label{figure-bou04}
\end{figure}

At the half filling, the density of Bethe roots satisfies
\begin{eqnarray}
  \tilde{\rho}_{04}(w) &=& -
  \frac{1}{2L}\frac{e^{-\frac{|w|}{2}}+e^{-(1-\xi)|w|}+e^{-\xi'|w|}+1}
  {1+e^{-|w|}}\nonumber\\
  &+& \frac{e^{-|w|}}{1+e^{-|w|}}.\label{energy-general-05}
\end{eqnarray}
From Eq.\eqref{energy-general-05}, we obtain the surface energy as
\begin{eqnarray}
\label{surface-energy-bou04}
  E_{b4} &=&-\int_{-\infty}^\infty\frac{e^{-|w|}}{2}
  \frac{e^{-\frac{|w|}{2}}+e^{-(1-\xi)|w|}+e^{-\xi'|w|}+1}{1+e^{-|w|}}dw
  \nonumber\\
  &&+\frac{1}{\xi-\xi^2}+\frac{1}{\xi'-\xi'^2}.
\end{eqnarray}
The surface energies of the system \eqref{Hamiltonian} with given boundary parameters in regime (iv) are shown in Fig.\ref{figure-bou04}, where the blue curves
are the results obtained by Eq.\eqref{surface-energy-bou04} and the red stars are the ones calculated by the DMRG.
From Fig.\ref{figure-bou04}, we see that the analytical result \eqref{surface-energy-bou04} agrees with the DMRG data.

\section{Conclusions}
\label{sec6}

In this paper, we have studied the physical quantities of one-dimensional supersymmetric $t-J$ model with unparallel boundary magnetic fields, which is a typical
$U(1)$-symmetry broken quantum integrable strongly correlated electron
system. At zero temperature, we find that the contribution of inhomogeneous term in the eigenvalue of transfer matrix can be neglected in the thermodynamic limit.
From the analysis of reduced Bethe ansatz equations, we obtain the patterns of Bethe roots. Based on them, we calculate the ground state energy and surface energy in different regimes of
boundary parameters. We also find that there exist some stable boundary bound states for certain boundary fields.

It is interesting to extend the present analysis to the finite temperature.
The reduced Bethe ansatz equations take the form of product, thus the thermodynamic Bethe ansatz \cite{takasha}
can be applied to calculate the thermodynamic properties such as elementary excitations,
spin-charge separation, free energy and magnetic susceptibility.
The analysis considered here also allows to study the corresponding conformal field theory in a geometry with boundary \cite{conformal}.
If the inhomogeneous term can not be neglected, then the $t-W$ scheme might be a good candidate to study the finite temperature thermodynamics for present model \cite{qiao}.
The exact finite size effect of the system (\ref{Hamiltonian}) starting from the inhomogeneous Bethe asnatz equations (\ref{bae_prime_01})-(\ref{bae_prime_02}) is also an
interesting topic.

\section*{Acknowledgments}

The financial supports from the National Natural Science Foundation of China
(Grant Nos. 12175180, 12104372, 12074410, 12047502, 11934015, 11975183 and 11947301), Major Basic Research Program of Natural Science of Shaanxi Province (Grant Nos. 2017KCT-12 and 2017ZDJC-32),  the Strategic Priority Research Program of the Chinese Academy of Sciences (Grant No. XDB33000000),
Australian Research Council (Grant No. DP 190101529), and Double First-Class University Construction Project of Northwest University are gratefully acknowledged.


\begin{thebibliography}{99}%

\bibitem{hubbard01} P. W. Anderson, The resonating valence bond state in $La_2CuO_4$ and supersonductivity, Science {\bf 235}, 1196 (1987).
\bibitem{zhang1988} F. Zhang and T. Rice, Effective Hamiltonian for the superconducting $Cu$ oxides, Phys. Rev. B {\bf 37}, 3759 (1988).
\bibitem{zeng02} P. B. Wiegmann, Superconductivity in strongly correlated electronic systems and confinement versus deconfinement phenomenon, Phys. Rev. Lett. {\bf 60}, 821 (1988).
\bibitem{hubbard02} F. H. L. Essler, V. E. Korepin and K. Schoutens, New exactly solvable model of strongly correlated electrons motivated by high $T_c$ superconductivity, Phys. Rev. Lett. {\bf 68}, 2960 (1992).
\bibitem{hubbard04} S. Reja, J. V. D. Brink and S. Nishimoto, Strongly enhanced superconductivity in coupled $t-J$ segments, Phys. Rev. Lett. {\bf 116}, 067002 (2016).

\bibitem{Lai} C. K. Lai, Lattice gas with nearest-neighbor interaction in one dimension with arbitrary statistics, J. Math. Phys. {\bf 15}, 1675 (1974).
\bibitem{zeng06} B. Sutherland, A general model for multicomponent quantum systems, Phys. Rev. B {\bf 12}, 3795 (1975).
\bibitem{sarkar1990} S. Sarkar, Bethe-ansatz solution of the $t-J$ model, J. Phys. A: Math. Gen. {\bf 23}, L409 (1990).

\bibitem{zeng03} D. F\"{o}rster, Staggered spin and statistics in the supersymmetric $t-J$ model, Phys. Rev. Lett. {\bf 63}, 2140 (1989).
\bibitem{essler199202} F. H. L. Essler and V. E. Korepin, Higher conservation laws and algebraic Bethe ansatz for the supersymmetric $t-J$ model, Phys. Rev. B {\bf 46}, 9147 (1992).
\bibitem{zeng04} A. Foerster and M. Karowski, Completeness of the Bethe states for the supersymmetric $t-J$ model, Phys. Rev. B {\bf 46}, 9234 (1992).
\bibitem{zeng05} A. Foerster and M. Karowski, Algebraic properties of the Bethe ansatz for an $spl(2,1)$ supersymmetric $t-J$ model, Nucl. Phys. B {\bf 396}, 611 (1993).
\bibitem{zeng08} A. Foerster, M. Karowski, The supersymmetric $t-J$ model with quantum group invariance, Nucl. Phys. B {\bf 408}, 512 (1993).


\bibitem{schlo1987} P. Schlottmann, Integrable narrow-band model with possible relevance to heavy-fermion systems, Phys. Rev. B {\bf 36}, 5177 (1987).
\bibitem{Bares-1991} P. A. Bares, G. Blatter and M. Ogata, Exact solution of the $t-J$ model in one dimensional at $2t=\pm J$: Ground state and excitation spectrum, Phys. Rev. B {\bf 44}, 130 (1991).
\bibitem{bares1990} P. A. Bares and G. Blatter, Supersymmetric $t-J$ model in one dimension: Separation of spin and charge, Phys. Rev. Lett. {\bf 64}, 2567 (1990).
\bibitem{1} N. Kawakami and S. K. Yang, Correlation functions in the one-dimensional $t-J$ model, Phys. Rev. Lett. {\bf 65}, 2309 (1990).
\bibitem{2} E. D. Williams, Thermodynamics and excitations of the supersymmetric $t-J$ model, Int. J. Mod. Phys. B {\bf 09}, 3607 (1995).
\bibitem{3} G. J\"{u}ttner, A. Kl\"{u}mper and J. Suzuki, Exact thermodynamics and Luttinger liquid properties of the integrable $t-J$ model, Nucl. Phys. B {\bf 487}, 650 (1997).
\bibitem{zeng10} J. Sirker and A. Kl\"{u}mper, Thermodynamics and crossover phenomena in the correlation lengths of the one-dimensional $t-J$ model, Phys. Rev. B {\bf 66}, 245102 (2002).


\bibitem{zeng01} E. K. Sklyanin, Boundary conditions for integrable quantum systems, J. Phys. A {\bf 21}, 2375 (1988).


\bibitem{zeng09} A. Gonz\'{a}lez-Ruiz, Integrable open-boundary conditions for the supersymmetric $t-J$ model the quantum-group-invariant case, Nucl. Phys. B {\bf 424}, 468 (1994).
\bibitem{essler1996} F. H. L. Essler, The supersymmetric $t-J$ model with a boundary, J. Phys. A: Math. Gen. {\bf 29}, 6183 (1996).
\bibitem{4} Y.-K. Zhou and M. T. Batchelor, Spin excitations in the integrable open quantum group invariant supersymmetric $t-J$ model, Nucl. Phys. B {\bf 490}, 576 (1997).
\bibitem{5} G. Bed\"{u}rftig and H. Frahm, Open $t-J$ chain with boundary impurities, J. Phys. A: Math. Gen. {\bf 32}, 4585 (1999).
\bibitem{zeng11} H. Fan H and M. Wadati, Integrable boundary impurities in the $t-J$ model with different gradings, Nucl. Phys. B {\bf 599}, 561 (2001).
\bibitem{zeng101} W. Galleas, Spectrum of the supersymmetric $t-J$ model with non-diagonal open boundaries, Nucl. Phys. B {\bf 777}, 352 (2007).
\bibitem{zeng108} A. S. Mishchenko and N. Nagaosa, Electron-phonon coupling and a polaron in the $t-J$ model: from the weak to the strong coupling regime, Phys. Rev. Lett. {\bf 93}, 036402 (2004).
\bibitem{hubbard03} Y. Q. Chong, V. Murg, V. E. Korepin and F. Verstraete, Nested algebraic Bethe ansatz for the supersymmetric $t-J$ model and tensor networks, Phys. Rev. B {\bf 91}, 195132 (2015).

\bibitem{Cao13} J. Cao, W.-L. Yang, K. Shi and Y. Wang, Off-diagonal Bethe ansatz and exact solution a topological spin ring, Phys. Rev. Lett. {\bf 111}, 137201 (2013).
\bibitem{wang2015} Y. Wang, W.-L. Yang, J. Cao and K. Shi, {\it Off-diagonal Bethe ansatz for exactly solvable models} (Spring Press, Berlin, 2015).
\bibitem{zhang2014} X. Zhang, J. Cao, W.-L. Yang, K. Shi and Y. Wang, Exact solution of the one-dimensional super-symmetric $t-J$ model with unparallel boundary fields, J. Stat. Mech. P04031 (2014).

\bibitem{takasha} M. Takahashi, {\it Thermodynamics of one-dimensional solvable models} (Cambridge University Press, Cambridge, 2005).

\bibitem{wen2018} F. Wen, Z.-Y. Yang, T. Yang, K. Hao, K. Cao and W.-L. Yang, Surface energy of the one-dimensional supersymmetric $t-J$ model with unparallel fields, JHEP {\bf 06}, 076 (2018).

\bibitem{conformal} J. L. Cardy, Conformal invariance and surface critical behavior, Nucl. Phys. B {\bf 240}, 514 (1984).

\bibitem{qiao} P. Lu, Y. Qiao, J. Cao, W.-L. Yang, K. Shi and Y. Wang, $T-W$ relation and free energy of the Heisenberg chain at a finite temperature, JHEP {\bf 07}, 133 (2021).

\end{thebibliography}

\end{document}